*Article*

# Fitting an Equation to Data Impartially

Chris Tofallis

Statistical Services and Consultancy Unit, University of Hertfordshire, College Lane, Hatfield AL10 9AB, UK; ormsresearch@gmail.com

**Abstract:** We consider the problem of fitting a relationship (e.g., a potential scientific law) to data involving multiple variables. Ordinary (least squares) regression is not suitable for this because the estimated relationship will differ according to which variable is chosen as being dependent, and the dependent variable is unrealistically assumed to be the only variable which has any measurement error (noise). We present a very general method for estimating a linear functional relationship between multiple noisy variables, which are treated impartially, i.e., no distinction between dependent and independent variables. The data are not assumed to follow any distribution, but all variables are treated as being equally reliable. Our approach extends the geometric mean functional relationship to multiple dimensions. This is especially useful with variables measured in different units, as it is naturally scale-invariant, whereas orthogonal regression is not. This is because our approach is not based on minimizing distances, but on the symmetric concept of correlation. The estimated coefficients are easily obtained from the covariances or correlations, and correspond to geometric means of associated least squares coefficients. The ease of calculation will hopefully allow widespread application of impartial fitting to estimate relationships in a neutral way.

**Keywords:** functional relationship; data fitting; errors in variables; linear regression; multivariate analysis; measurement error model

**MSC:** 62J05

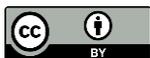

## 1. Introduction and Motivation

*"The regression line paradox. Regression lines are like one-way streets: you can only move in the authorized direction. If you want to reverse direction you must move across to the other regression line".* ([1], pp. 133–134)

An essential aspect of science is the extraction of knowledge from data. The natural sciences developed by collecting data from experiments (e.g., physics), or observation (e.g., astronomy), and then identifying a relationship connecting the measured variables. The resulting equations could be re-arranged to change the subject of the formula, as required. Later, the social sciences (e.g., economics, psychology) tried to emulate this approach. In statistics courses, the tool employed to estimate the relationship makes the (often unstated) assumption that all but one of the variables does not contain any measurement error. This tool is regression—it requires the user to select a dependent variable (which contains all the noise in the system) and which is treated differently from the other variables. Consequently, regression models cannot be re-arranged to change the subject of the equation. Our interest in this paper is to establish a simple method for estimating a relationship without having to select one variable as being special in this way. Rather, we seek to treat all variables in the same way—i.e., an impartial or symmetric approach. Orthogonal regression [2] is sometimes suggested for this problem; this applies least squares to the perpendicular distances to the fitted line or (hyper)plane. This involves summing squares of ALL the variables. However, changing the measurement units of any variable leads to an equation that is not equivalent to the original one. "It suffers from the disadvantage that … it is not invariant under change of scale" ([3], p. 31), i.e., this approach is not units-invariant (or scale-invariant). Serious difficulties arise when measuring distance in a space where different directions are measured in different units. Consider three points defining a right-angled triangle in a coordinate space where the base is in kg units and the height is in metres; what units does the hypotenuse have? There are of course various ways of making data dimensionless, e.g., dividing by the mean, dividing by standard deviation, taking logs, etc. However, the resulting equation from each of these will differ, depending on the (arbitrary) choice of normalization used. As Blankmeyer observes [4]: "this method has a major disadvantage: the coefficients in an orthogonal regression are not equivariant; they change in a complicated way when a variable is rescaled".

We avoid this issue by basing our approach on the symmetric concept of correlation. Correlation has the benefit of not being affected by measurement units. It also does not require the specification of a direction in which to minimize distances (least squares).

We shall consider the problem of finding a single underlying linear relationship between multiple variables. We wish to treat all variables on the same basis and so we make no distinction between dependent and independent variables: technically, we are seeking a functional relationship.

The Cambridge Dictionary of Statistics [5] defines it as: 'The relationship between the 'true' values of variables, i.e., the values assuming that the variables were measured without error'. Lindley [6] says 'a functional relationship is required for the statement of laws in the empirical sciences which would hold if no errors existed'. Kendall and Stuart [7] introduce the subject thus: 'It is common in the natural sciences, and to some extent in the social sciences, to set up a model of a system in which certain mathematical (not random) variables are functionally related.' They stress the difference between a functional relationship and regression: "One consequence of the distinctions we have been making has frequently puzzled scientists. The investigator who is looking for a unique linear relationship between variables cannot accept two different lines … Our discussion should have made it clear that a regression line does not purport to represent a functional relation between mathematical variables or a structural relation between random variables." (p. 394). They later give an example dealing with the pressure, volume and temperature of a gas, i.e., the gas law in physics. It is assumed that measurement error (or noise) is associated with all the variables. A simpler example, often taught in school science, is the relationship between the extension of a wire or spring and the amount of weight attached to it (Hooke's law). One would like the relationship to give the extension for a given weight, and conversely to give the same weight value for that extension, i.e., a one-to-one and invertible function. Ordinary regression cannot provide this. If knowledge regarding the relative magnitudes of the measurement errors were available then we could apply the theory of measurement error models, also known as errors-in-variables models (see the books by Cheng and Van Ness, [8], or Fuller, [9]). However, we shall assume that error variances from replicated measurements are not available. Our main assumption will be that all variables are measured with equal reliability. Equal reliability (a dimensionless value between 0 and 100%) is not the same as equal error variance—the latter possibly not being meaningful when variables are measured in different units. Furthermore, we shall not make any assumptions regarding the distributions of the variables—which is the field of 'structural relationships'.

We now explain why ordinary least squares (OLS) regression is not appropriate for our task. For the case of two variables, x and y, the linear regression of y on x leads to a line whose equation is:

$$y = r(s_y/s_x)x + c,$$

where r is the sample correlation and s the sample standard deviation.

If we assume that the data have been standardized by subtracting the mean and dividing by the standard deviation, this equation simplifies to $y = rx$. Now consider the regression of x on y; this leads to the relationship $x = ry$, or $y = x/r$.

These two lines do not agree with each other; see Figure 1, which is taken from Pearson's 1901 paper [2], where he recognised that the 'line of best fit' lies between the two regression lines. The only point of agreement between the regression lines is where they intersect, which corresponds to the origin for standardized data, and which is the centroid, or the point of means, for the original data. Lindley [10] described the problem in this way:

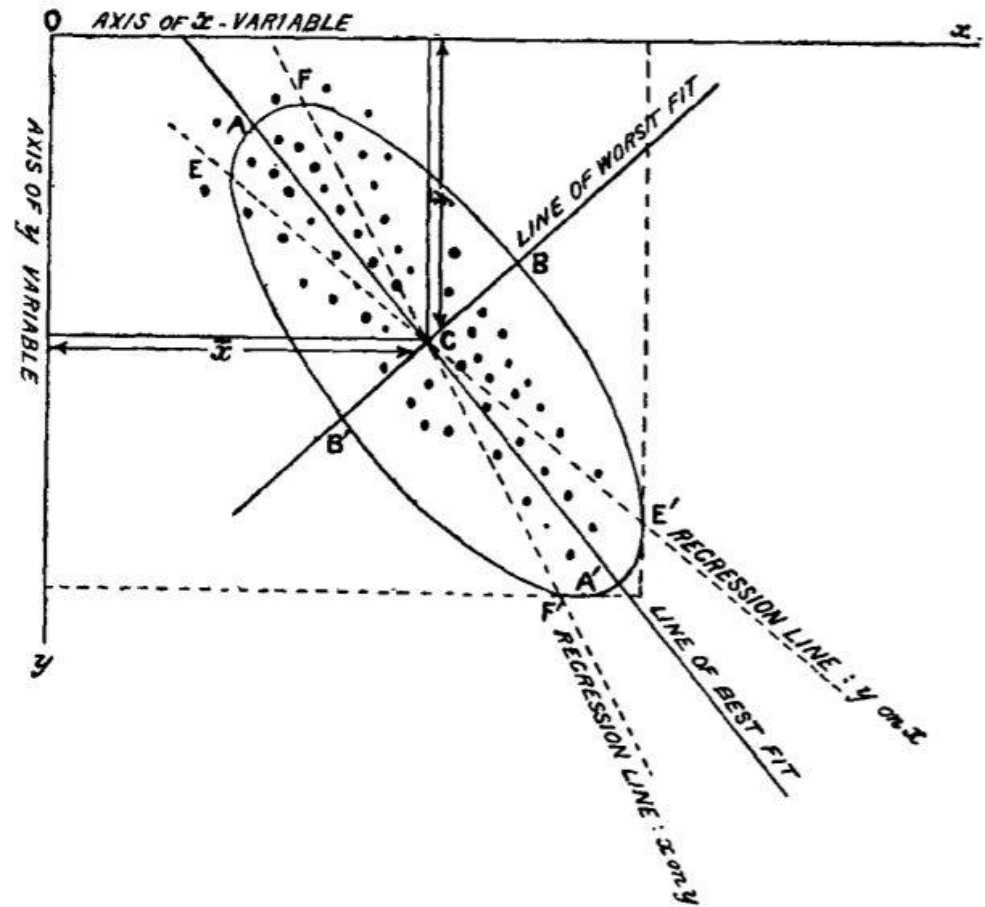

**Figure 1.** Comparison of the ordinary regression, reverse regression, and 'line of best fit'. Source: [2]. Ordinary regression is appropriate if the x values have zero error. Reverse regression is appropriate if the y values have zero error. But, in general, the relationship will lie between these extremes.

"It is known that if there is a linear functional relationship between two quantities, if the only observations available on these quantities are subject to error… then there is no satisfactory estimate of the functional relationship unless the relative accuracy of measurement of the two quantities, or some similar knowledge, is available. Furthermore, in this situation there are three relationships which exist between the two quantities, namely, the original functional relationship and two regression equations which, in this case, will be linear. The experimenter who encounters this situation is therefore faced with a difficulty and a complexity in the mathematics which he feels is foreign to his practical situation, and it is not therefore surprising that numerous attempts have been made to remove the undesirable features of the problem".

Thus, if we are trying to describe a law-like relationship from the data, OLS unfortunately provides two inconsistent relations. Suppose we wanted to estimate the 'slope' (the term by which we will refer to the rate of change of y with x in what follows, including the multivariate case). One application of the slope might be to predict the extent of a change: in this case, the change in y for a given change in x. The ordinary regression estimates the slope as being equal to r, whereas the reverse (least squares) regression estimates this as 1/r (using the same axes). Thus if, say, the correlation were 0.71, then one slope estimate would be twice as large as the other, and hence the estimated changes would also differ by a factor of two! This is not a difference that can be ignored.

## 2. Method for Two Variables

Let us begin by imagining a situation where, in the absence of measurement error, all data points fit a straight line perfectly. The correlation between the two variables would be perfect r = ±1. The introduction of measurement error will make the correlation change from its perfect value. We would hope or prefer that the level of noise in the data would not affect our estimate of the underlying relationship. Sadly, this is not the case: any measurement error associated with the x-variable will cause the y on x regression slope to be biased downward in magnitude, i.e., closer to zero. This effect is known as regression dilution, or regression attenuation. The slope estimate keeps falling as the noise increases (and the correlation r moves toward zero). By contrast, we notice that the reverse OLS slope estimate using the same axes, (change in y)/(change in x), will increase in magnitude as the noise in y rises. Thus, the two regression lines diverge as the noise increases (and therefore r decreases). Conversely, as the noise declines, the two lines converge; see Figure 2.

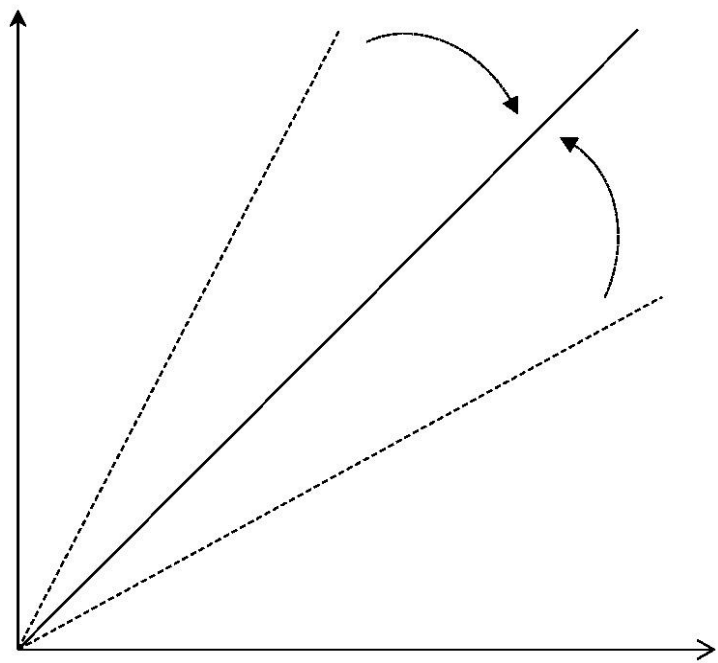

**Figure 2.** The perfect correlation relationship. Assuming there exists an underlying linear relationship, any noise in the data will cause the ordinary least squares regression line to under-estimate the slope (lower dashed line). The reverse regression line over-estimates the slope (upper dashed line). If measurement error in the data is reduced, the correlation improves, and the two lines converge to the perfect correlation line.

In seeking a single equation to summarize the data we are therefore faced with a situation where we have two lines of differing slopes which intersect at the centroid (point of means), neither of which is acceptable. Tan and Iglewicz [11] point out that 'the word "regression" implies the asymmetric roles of the dependent (response) variable and the independent (regressor) variable'. In fact, what we are seeking is not a regression, but a relationship that treats all variables symmetrically or impartially.

Clearly, if both variables contain errors, then the 'true line', however defined, lies somewhere in between the two regression lines, and we now develop an argument for how to estimate it.

Let us assume that the data are affected by noise (possibly due to measurement error) in the data, and that there is a true underlying linear relationship, perhaps due to a physical or other scientific law. Now imagine 'turning down the noise' (e.g., by using more accurate instruments) so that the scatter is reduced and the correlation rises. Then, the above two OLS lines (y = rx and y = x/r) will gradually converge, and eventually coincide when the correlation is perfect (zero noise); see Figure 2. In standardized variables, the equation of the resulting unique line will take the form:

y = ±x (where the sign is given by the correlation).

In terms of the original unstandardized data, the slope is the ratio of standard deviations $\pm(s_y/s_x)$, which has a magnitude of unity on standardization.

Because of the way we have arrived at this line, we shall refer to it as the perfect correlation line. We choose this name because it also provides the key to unlocking the estimation procedure in higher dimensions. We shall imagine a correlation 'dial' as a way of adjusting the noise in the data such that when the noise is zero (perfect correlation), we uncover the underlying functional relationship.

This line is known as the geometric mean functional relationship (GMFR), see, e.g., [12] or [13], because the slope is the geometric mean of the two OLS slopes. (It is also known as the reduced major axis [14]). Notice that by taking the geometric mean of the slopes, the presence of the correlation in the slope formulae is removed. (This would not occur if the arithmetic mean were used.) This is an important and valuable property, because, in general, we would not want noise to affect the slope estimate or underlying relationship in a systematic way, as this would lead to bias. Viewing the (complement of the) correlation as an indicator of noise level in the data, it seems appropriate that the correlation should not be present in the slope estimation formula for a situation where it is believed that a genuine functional relationship exists.

Viewed in another way, any researcher who regresses y on x and is later able to obtain more accurate (less noisy) data will always conclude that their previous slope estimate was too low, whereas any researcher who regresses x on y will later find that their original slope estimate was too high, and that more accurate data provided a lower estimate. It is therefore natural to follow the above path of taking a line intermediate between these two estimators, which is both convenient and natural and is scale invariant. The perfect correlation line provides this by using the geometric mean of the slopes.

This line has a number of interesting and attractive properties; ref. [15] summarises nine of them. For example, ref. [16] proves that for a large class of distributions, including the bivariate normal; this line pairs off X and Y values such that the proportion of x-values below X is the same as the proportion of y-values below Y. In other words, it pairs off values with equal percentiles. One application for this is when matching scores on two tests which are designed to measure the same aspect of performance. Another application is to relate measurements of the same quantity from two different instruments. By definition, the OLS line has the least sum of squared residuals in the y-direction. Similarly, no other line has a lower sum of squares in the x-direction than the reverse OLS line. In general, no line can simultaneously achieve both of these optimal properties. Greenall [16] proved that the perfect correlation line (GMFR) increases both sums of squares by the same factor or percentage.

A geometric property of the GMFR leads to it also being referred to as the 'least areas line': if one constructs triangles by joining each data point to the fitted line using horizontal and vertical segments, then the fitted line minimizes the sum of these triangular areas. Changing the scale of one of the variables merely stretches or compresses in that direction and the triangular areas scale accordingly. This leads to the useful property of scale-invariance for the fitted line. This is particularly valuable when variables are measured in different units, and is a property that orthogonal regression does not possess. The least areas property implies that the sum of products of the vertical and horizontal deviations to the line is also minimized, hence the name 'least products line'. This is also equivalent to minimizing the sum (over all data points) of the geometric means of squared deviations to the line in both dimensions.

## 3. Treating Variables in the Same Way: Equal Reliability

When measurement error affects both variables, it is possible to derive a maximum likelihood estimate of the slope under the assumption of a bivariate normal distribution. Unfortunately, this requires knowledge of the variance of the errors [17]. Such information can be obtained if one has the luxury of replicated observations, which is typically not the case. Nevertheless, it is of interest to ask: under what circumstances does the above GMFR line provide the maximum likelihood solution?

If we express the observed values as $x = x_{true} + e_x$ and assuming the measurement errors ($e_x$) are uncorrelated with the true values, we expect:

$$var(x) = var(x_{true}) + var(e_x)$$

Note that since the mean error is zero, $var(e_x)$ is simply the mean square measurement error.

The reliability (coefficient) for x is defined by:

$$Rel(x) = var(x_{true})/var(x), \text{ i.e., } [var(x) - var(e_x)]/var(x)$$

or $1 - [var(e_x)/var(x)]$

Thus, as the error variance tends to zero, the reliability tends to its maximum value of unity.

It is well established that the maximum likelihood estimator for the slope corresponds to the geometric mean slope when the ratio of the error variances is equal to the ratio of the total variances (see, for example, [18]):

$$var(e_y)/var(e_x) = var(y)/var(x)$$

$$\text{i.e., } var(e_x)/var(x) = var(e_y)/var(y)$$

$$\text{so } Rel(x) = Rel(y)$$

Thus, the geometric mean slope is the same as the slope estimated by maximum likelihood when both variables are measured with equal reliability. Notice that changing the units of measurement does not affect the reliability. This makes this a more valuable approach than orthogonal regression, which lack units-invariance, and which is maximum likelihood for equal error variances—an assumption which is less realistic. According to [8] (p. 81) and [19] (p. 287), there are many applications, e.g., in econometrics, where the equal variance assumption is not useful. Our approach is to shift focus away from variance toward reliability: in the absence of information on the relative reliability of the data, it seems natural to assume equal reliability. This may be viewed as a heuristic approach.

We can, in fact, derive the geometric mean slope expression from the equal reliability property in a distribution-free way: As before, let the observed value be $y = y_{true} + e_y$.

If there is a linear relationship between the true values of y and x, with slope $\beta$, it follows that

$$var(y_{true}) = \beta^2 \, var(x_{true})$$

and if both variables have the same reliability, $\lambda$, then

$$\lambda \, var(y) = \beta^2 \, \lambda \, var(x)$$

from which it follows the slope can be estimated from $b^2 = var(y)/var(x)$.

Thus, b is the geometric mean of the two least squares slopes, i.e., the GMFR slope.

McArdle [18] summarises the situation thus: "If the two variables are highly correlated, then there is unlikely to be a problem—the OLS of y on x is $r^2$ times the OLS of x on y. All available solutions lie between these two extremes. If, however, the range of possible slopes is too large for the purposes of the study, then more information will be needed before it can be improved".

How the uncertainty associated with the location of the data points is decomposed between x and y determines how close to either limit the true line is located.

In the absence of any information as to the relative reliability of the data variables, it seems reasonable to treat them on the same basis, and so we advocate the geometric mean line we have been discussing.

## 4. Generalization to Multiple Dimensions

We now extend the above perfect correlation or geometric mean approach to estimate a functional relationship involving *multiple* variables:

$$\Sigma \beta_j X_j = \text{constant}$$

where $X_j$ represent the 'true' values measured without error, and we shall estimate the $\beta_j$ by $b_j$.

Previous attempts at generalisation have focused on the least areas geometric property of the geometric mean functional relationship in two dimensions. Draper and Yang [20] generalized this aspect in a particular way. They considered the geometric mean of the distances to the (hyper)plane in each dimension and applied least squares to that quantity. Unfortunately, the resulting optimization problem is nonlinear and no expression is available for the coefficients. They proved that in the space of coefficients the solution lies in the simplex defined by all the least squares solutions (taking each variable in turn as the dependent variable).

A different generalization was developed by Tofallis [21,22], which was to view the least areas in the two-dimensional case as a minimization of the sum of products of distances to the line in each dimension. Thus, in three dimensions, the problem becomes one of minimizing the sum of volumes of the tetrahedra created by each data point and the plane. Once again, this involved nonlinear optimization. In a doctoral thesis, Li [23] explored this approach by carrying out simulations to investigate the distribution of coefficient estimates and bootstrapping to obtain confidence intervals. The effects of sample size, sampling error and correlation among variables on the estimates were also studied. Another variation is the least sum of geometric mean deviations [15]. This leads to a linear objective function with all constraints linear, apart from one which sets the product of coefficients to unity. In the following sections, we show how to fit a functional relationship without having to solve any nonlinear optimization problems.

*4.1. Three Variables: Fitting a Plane*

We now describe how to fit a plane to data on three variables, x, y, z, in a symmetric fashion. (This section can be skipped by those who wish to go directly to the general result for multiple variables.) To begin with, imagine the data points are scattered about but lying perfectly on a plane. If we view a scatterplot of all data points in the x-y plane, we will not see a straight line because the points have different values of z. Hence, we cannot assume conditions saying that the bivariate correlations should be perfect, as we did before. However, if we only plot points with the same z-value, we will see evidence of a straight line because we have taken a cross-section through the plane. The way forward in our derivation is to begin by imagining that the correlation between x and y for fixed z, be perfect. We denote this partial correlation by $r_{xy.z}$. Naturally, the perfect correlation condition also applies to the other two partial correlations, and this will permit us to deduce the equation of the plane being fitted to the data.

The theory of *partial correlation* was initially developed by Yule [24], where he defines the partial correlation between x and y, keeping z constant, as:

$$r_{xy.z} = (b_{xy} \, b_{yx})^{½} \qquad (1)$$

where $b_{yx}$ is the OLS coefficient of x when y is regressed on x and z, and $b_{xy}$ is the coefficient of y when x is regressed on y and z. Yule used the older notation $b_{xy.z}$ to emphasize that it was a partial coefficient, i.e., keeping z constant, but in modern notation the OLS coefficients are in fact partials.

In one regression, the rate of change of y with x for fixed z is estimated by $b_{yx}$, whereas in the other regression, it is estimated by $1/b_{xy}$. As the noise is reduced to zero, the partial correlation tends to unity and, in the limit, we have from (1):

$$b_{yx} = 1/b_{xy} \qquad (2)$$

This demonstrates that the two least squares slopes will only agree when there is perfect partial correlation.

Multiplying (2) by $b_{yx}$, we obtain

$$b^2_{yx} = b_{yx}/b_{xy} \qquad (3)$$

Denoting the resulting value for the 'perfect slope' in the x-y plane by $b^*_{yx}$ gives us

$$b^*_{yx} = (b_{yx}/b_{xy})^{½} \qquad (4)$$

The sign of this square root is the sign of either $b_{yx}$ or $b_{xy}$; both of these quantities will always have the same sign, as the left side of (3) is positive.

Note that the reciprocal of $b_{xy}$ refers to the rate of change of y with x. Thus, (4) shows that our perfect correlation 'slope' is the geometric mean of the two OLS slopes. Fortunately, and very conveniently, this is the same fundamental connection as in the two-dimensional case.

We next show that these 'perfect' regression coefficients can be expressed in terms of simple correlations. The textbook of Spiegel ([25], p. 270, Equation (5)) shows that OLS standardized regression coefficients can be expressed as follows:

$$b_{yx} = (r_{yx} - r_{yz} r_{xz})/(1 - r^2_{xz})$$

$$b_{xy} = (r_{yx} - r_{xz} r_{yz})/(1 - r^2_{yz})$$

Substituting these into (4) gives:

$$b^*_{yx} = [(1 - r^2_{yz})/(1 - r^2_{xz})]^{½}$$

When taking the square root, the sign of this partial coefficient should match the sign of the associated partial correlation, $r_{yx.z}$.

Similarly, the rate of change of y with z is estimated by

$$b^*_{yz} = [(1 - r^2_{yx})/(1 - r^2_{xz})]^{½}$$

Putting these together implies that, in terms of standardized variables, the equation of the fitted plane can be written very elegantly as:

$$x(1 - r^2_{yz})^{½} + y(1 - r^2_{xz})^{½} + z(1 - r^2_{yx})^{½} = 0$$

*4.2. The General Case: Multiple Variables*

4.2.1. Deducing the Perfect Correlation Coefficients

In this section, we show how to obtain the coefficients in the perfect correlation functional relationship for any number of variables; this is our estimate of the true relationship. Whereas in two dimensions, the key step was to let the correlation magnitude tend to unity, we will now let the *partial* correlations play the same role. A remarkable, but little-known, fact is that one can compute all possible linear regressions simultaneously (without having to take each variable in turn as the dependent variable). This is because the standardized coefficients for all regressions are provided by the inverse of the correlation matrix—also known as the precision matrix or concentration matrix. (Inverse matrices can be computed using the standard Excel matrix function MINVERSE. They can also be found using statistical software such as R, SAS, etc.) The correlation matrix is symmetric, its inverse is also symmetric, and the ith row or column gives the regression coefficients for the symmetric form $\Sigma b_j x_j = 0$, when $x_i$ is the dependent variable. See, for example, [26] (Equation (3)), or [27] (Section 4.2).

Next, we show how one can obtain the regular (non-standardized) OLS coefficients using the inverse of the covariance matrix instead of the inverse of the correlation matrix.

Denote the inverse covariance matrix by $K^{-1}$ and its elements by $c_{ij}$. The partial correlations are related to the $c_{ij}$ as follows (see [28], or [29], p. 13 and p. 125):

$$r_{ij} = c_{ij}/(c_{ii} c_{jj})^{1/2}$$

where the left side is the partial correlation between variables $x_i$ and $x_j$ when the remaining variables are held constant.

As the noise tends to zero, the partial correlations will tend to unity in magnitude, so that in the limit, the above relation leads to $c_{ij}^2 = c_{ii} c_{jj}$. This means that all the off-diagonal elements of $K^{-1}$ can be expressed in terms of the diagonal elements alone, as the noise approaches zero.

The first row of $K^{-1}$ will then be $[c_{11}, (c_{11}c_{22})^{1/2}, (c_{11}c_{33})^{1/2}, \ldots]$, and the second row will be $[(c_{11}c_{22})^{1/2}, c_{22}, (c_{22}c_{33})^{1/2}, \ldots]$, etc.

Also, when the noise tends to zero, all the OLS regressions will converge so that the rows of $K^{-1}$ will be equivalent (up to a multiple); this means that any of these rows will provide the coefficients we need. For example, taking the first row and dividing through by $(c_{11})^{1/2}$ gives the jth coefficient as

$$b_j = (c_{jj})^{1/2}$$

The same result arises by dividing through the second row by $(c_{22})^{1/2}$.

In summary, the perfect correlation functional relationship is represented in symmetric form by

$$\Sigma (c_{jj})^{1/2} x_j = \text{constant}$$

where the constant is found by substituting the means of the variables, and $c_{jj}$ are the diagonal elements of the inverse covariance matrix. This is the key result of this paper.

4.2.2. Connection with the Geometric Mean of OLS Coefficients

We now show how the geometric mean slopes property in two dimensions extends to higher dimensions.

When there are multiple (*p*) variables, Yule [24] gives the general relationship between partial correlations and OLS coefficients (b), which generalizes (1) above:

$$r_{12 \cdot 3 \ldots p} = (b_{12} b_{21})^{1/2}$$

where the left side is the correlation between variables $x_1$ and $x_2$ when the remaining variables are held constant.

Once again, we consider reducing the noise so that the partial correlations tend to unity. In the limit, the above expression leads to $b_{12} = 1/b_{21}$, so the rate of change of $x_1$ with $x_2$, ($b_{12}$), will be equal to $1/b_{21}$, and the partial slopes agree when we regress $x_1$ on $x_2$, and $x_2$ on $x_1$. Multiplying both sides by $b_{12}$ implies $b_{12} = (b_{12}/b_{21})^{1/2}$

Let us denote this value by $b^*_{12}$.

Once again, we have the geometric mean of the two associated OLS rates of change. The other $b^*$ coefficients can be found similarly:

$$b^*_{ij} = (b_{ij}/b_{ji})^{1/2} \qquad (5)$$

This shows that the (partial) rate of change of $x_i$ with $x_j$ is the geometric mean of the estimates of this quantity from the OLS regressions using $x_i$ and $x_j$ as dependent variables.

Hence, the property which gives the name to the geometric functional relationship in two dimensions is now seen to persist in higher dimensions.

The fact that the required coefficients are directly expressible in terms of OLS coefficients brings with it the valuable property of uniqueness: provided we have more observations than parameters, and assuming there is no degeneracy in the data, then the fitted relationship will be unique.

4.2.3. Connection with Standard Errors and Coefficients of Determination

Kendall and Stuart [7] (p. 338) show that an OLS regression coefficient can be expressed as the ratio of a conditional covariance to a conditional variance:

$$b_{ij} = s_{ij.k}/s^2_{j.k}$$

where the suffix notation indicates that all variables apart from those on the left of the point are held fixed. Thus, $s^2_{j.k}$ is the variance which remains unique to the jth variable after accounting for its variation by all the other variables. Substituting this (and similarly for $b_{ji}$) into our geometric mean estimator (5), we obtain:

$$b^*_i = \pm(s_{i.k}/s_{j.k}) \qquad (6)$$

Note that this generalizes the bivariate case where the slope was $\pm(s_y/s_x)$, the only difference is that in three or more dimensions we have to use *conditional* standard deviations.

Thus, in the symmetric form of the fitted equation $\Sigma b^*_j x_j = 0$, the coefficient magnitudes are $1/s_{j.k}$, i.e., reciprocals of the *conditional standard deviations* (also known as residual standard errors or root mean square errors). This implies that the product of the coefficient with the residual standard error is the same (magnitude 1) in each dimension.

$s^2_{j.k}$ is also referred to as the unexplained variance, and for standardized data $s^2_{j.k} = 1 - R_j^2$ where $R_j^2$ is the coefficient of determination when $x_j$ is regressed on the other variables. (The fraction of the variance in $x_j$ is explained by the other variables).

Ref. [26] relates $1 - R_j^2$ to the jth diagonal element of the inverse of the correlation matrix (see also ([7], p. 347) thus:

$$1/(1 - R_j^2) = R^{-1}_{jj}$$

where the right side denotes the diagonal elements of the inverse of the correlation matrix. These again provide the required coefficients. If the data are not standardized, one uses the inverse covariance matrix instead.

Note that we do not require computation of eigenvectors or eigenvalues. This ease of calculation will hopefully allow more widespread application of impartial fitting to estimate relationships.

## 5. Scale-Invariance and Orthogonal Regression

If we re-scale by changing the measurement units of a variable, we would expect the associated coefficient to adjust accordingly. For example, if we switch the units of a variable from metres to kilometres, then the coefficient of that variable should become a thousand times larger, so that their product stays the same. This valuable feature is present in OLS regression. Since the coefficients in the proposed method can be directly obtained as geometric means of OLS coefficients, it follows that the equation relating all the variables will be scale-invariant. Specifically, from (5), suppose a change of units causes an OLS coefficient to change by a factor $\alpha$, then $b_{ij}$ becomes $\alpha b_{ij}$ and the reverse regression slope $1/b_{ji}$ becomes $\alpha/b_{ji}$. From (5), the effect on the geometric mean slope will be that the right-hand side becomes $(\alpha^2 b_{ij}/b_{ji})^{1/2} = \alpha\, b^*_{ij}$. Hence, it naturally scales as one would require.

When fitting equations by treating all variables on the same basis, the informal term 'symmetric regression' is sometimes used [30]. One symmetric approach that is often proposed is orthogonal regression, also known as total least squares. This minimizes the sum of squares of perpendicular distances to the fitted line or plane. While this can produce a single equation relating all variables, and appears not to treat any variable preferentially, it is well-known that it is not scale invariant [3], and so does not satisfy the above property. As explained in our introduction, the underlying reason is that distance is not a well-defined quantity in a space where each dimension is measured in different units. It is therefore not an attractive approach unless all variables happen to be measured in the same units. It is also more difficult computationally compared to the method presented here.

Taagepera [31] is a strong proponent of symmetric regression, devoting a chapter to it in his book *Making Social Sciences More Scientific*. He argues that OLS models "cannot form a system of interlocking models, because they are not unique, cannot be reversed, and lack transitivity. Scale-independent symmetric regression avoids these problems by offering a single, reversible, and transitive equation". Reversibility requires that if $y = f(x)$ then $x = f^{-1}(y)$, i.e., we should be able to rearrange the resulting relationship just as we do with any algebraic equation or scientific law. By contrast, OLS models are unidirectional. He points out that if one were testing a well-established linear law, the OLS slope would always be shallower than the true slope. Moreover, this unwanted artefact would arise whichever variable one plotted on the horizontal axis!

In two dimensions, he proposes the geometric mean functional relationship as the 'only scale-independent symmetric regression' that avoids all of the above difficulties of OLS. As we have seen, this line has slope $s_y/s_x$, whilst for OLS, it is $rs_y/s_x$, which is unfortunately a mixed measure confounding (lack of) noise and slope. Taagepera illustrates the difficulty by posing the question: how much does bodyweight increase with height (i.e., what is the slope)? With OLS, the answer depends on how accurate the measuring instruments are—not in a random way, but in a systematic way: as the accuracy rises, so does the rate of increase! In Figure 2, this corresponds to the lower dashed line moving towards the full line as accuracy rises. This is obviously not a real effect, but an unwanted artefact of conventional least squares models.

## 6. Numerical Illustration

To demonstrate the calculations, we use an example discussed in a book on measurement error models ([8], p. 142). The 36 data points were generated as a 6 by 6 square lattice in the $x_1$-$x_2$ plane (mean of $x_1 = 0.9$, mean of $x_2 = 0.88$), with y-values calculated from $y = 1 + 2x_1 + 3x_2$. This constituted the true underlying equation to be estimated. Measurement error noise was added to each of the three variables, distributed normally with zero mean and unit variance. (We are grateful to the authors for pointing out the errata in some signs in their published data.)

The covariance matrix (with y treated as the third variable) for the noisy data was:

| 9.54 | 1.29 | 17.23 |
|---|---|---|
| 1.29 | 9.40 | 28.14 |
| 17.23 | 28.14 | 112.88 |

Ref. [8] reported that ordinary regression gives $y = $ constant $+ 1.43\ x_1 + 2.80\ x_2$, which is very different from the underlying model. They then applied the relevant maximum likelihood model, which, in this case, corresponds to orthogonal regression.

This gave $y = $ constant $+ 1.52\ x_1 + 3.05\ x_2$, so the coefficients are closer to the true values of 2 and 3, and were described as 'quite good values'.

To apply our approach, we calculate the inverse of the above matrix, giving:

| 0.318 | 0.399 | −0.148 |
|---|---|---|
| 0.399 | 0.920 | −0.290 |
| −0.148 | −0.290 | 0.104 |

The coefficient magnitudes are the square roots of the diagonal elements, and their signs are provided by the signs in any row or column. This gives $0.564x_1 + 0.959x_2 − 0.322y = $ constant.

Dividing through by 0.322 to make the y-coefficient unity gives y = constant + 1.75 $x_1$ + 2.98 $x_2$.

These coefficients are closer to the true values of 2 and 3 than both the OLS and orthogonal regression estimates. This is particularly impressive, given that the variables do not actually have equal reliability—the y-values have a reliability 10 percentage points higher than the x-values. This gives an initial indication that the method has some robustness regarding the reliability assumption.

When our estimated functional equation is used for prediction in each of the three dimensions, the mean residual is zero, and variance of the residuals is, 3.62, 1.25 and 11.10 (for $x_1$, $x_2$ and y, respectively). Taking the reciprocal square roots of these variances gives the same coefficients, within a scaling factor. Thus, the product of the coefficient and the residual standard error is the same for each variable, as predicted in Section 4.2.3.

(For data sets with many variables, one may wish to avoid calculating the entire inverse matrix. One can, in fact, efficiently compute the diagonal elements alone. This is due to the positive definiteness property of covariance and correlation matrices, see [32].)

## 7. Discussion

*"In the past many people, mostly non-statisticians, have used regression techniques in the belief that they were thereby establishing natural laws. When statisticians have pointed out that they were doing nothing of the sort, the experimenters have reasonably asked what techniques they ought to have been using, and here there has sometimes been a quick change of subject"* (Pearce in the discussion of [33] p. 293).

Our purpose was to develop a method for describing the linear structure in data on multiple variables when no information is available regarding relative noise levels (measurement errors), and one wishes to treat each variable in the same way, i.e., as being equally reliable. For the two variable case, Ehrenberg [34] (p. 240) points out that "Fitting a line by minimising the perpendicular deviations might seem an intuitively attractive approach, but it has a crippling disadvantage. Changing the units of one variable (as from inches to feet) results in a completely different equation". In fact, "The position of the major axis [orthogonal regression line] can be made to lie anywhere between the two predictive regressions, simply by changing the units of measurement. Since measurement units can be chosen quite arbitrarily, obviously the major axis is not a good statistic for characterizing the trend" [35] (p. 415). We therefore rejected orthogonal regression because it "is not invariant under a change of scale and is therefore open to the analyst's subjectivity unless the variables measure quantities in the same units" [36].

In our proposed approach, the variables may be measured in different units—the resulting symmetric functional relationship is both units-invariant and unique. The parameters in the formula can be estimated directly from the covariances. The key result is that the (partial) rates of change ('slopes' in each $x_i$-$x_j$ plane) are the geometric means of these slopes as estimated by the two OLS regressions using $x_i$ and $x_j$ as dependent variables. Hence, we have generalised the geometric mean functional relationship to multiple dimensions. Equivalently, and more directly, the coefficients can be found from the diagonal elements of the inverse of the covariance matrix.

In summary, the steps are:

- Compute the inverse of the covariance matrix;
- The square roots of the diagonal elements provide the magnitude of the coefficients;
- The signs of the coefficients are provided by any row or column of the inverse matrix;
- The constant in the equation is found by using the fact that the plane or hyperplane passes through the point of means of all the variables.

There are many aspects of the method that are now open and ripe for exploring. For example, which of the many elegant properties in the bivariate case carry over to the multiple variable case? Since we have made no distributional assumptions, there is no associated inference theory. However, since the coefficients are geometric means of pairs of OLS coefficients, these provide upper and lower bounds. One might also suggest using the intersection of the traditional OLS intervals for these pairs, i.e., the overlap of their confidence intervals. Resampling methods can also be used to obtain confidence intervals for the coefficients.

Although practically all measurements involve some form of error, the multivariable methods applied in analysing such data rarely take account of this important fact. According to [8] (p. 13) "one of the main reasons why standard regression is so much more common than measurement error modelling is that it is computably and analytically simpler than measurement error modelling". A recent survey [37] notes that "measurement error methods are not used frequently in the situations that merit their use. There are a couple of reasons which hinder the use of measurement error methodology. Lack of adequate understanding of measurement error impacts is perhaps the primary reason. Developing proper methods to address measurement error effects requires analytical skills and knowledge of the measurement error models. This presents a significant hurdle for introducing suitable corrections in the analysis. Furthermore, general-purpose software packages are lacking for analysts to implement the available correction methods".

This paper was inspired by the work of Draper and Yang [20], which also appeared in the well-known book by Draper and Smith [12]. In concluding their paper, they stated that their computational approach "provides a practical solution to a difficult problem". We have managed to avoid their computationally intensive estimation process and have expressed the coefficients in terms of established statistics. We hope that the above analytic solution provides a significant advance such that users have access to a simple and practical tool for a difficult problem. The ease of calculation will hopefully allow more widespread application of symmetric fitting to estimate functional relationships impartially. The beauty of the approach presented here is that by treating all variables in the same way, we arrive at a very elegant result which does not require specialist software, and can be taught in data analysis at introductory level.

**Author Contribution:** An early version of this work appeared as the following Working Paper: Tofallis, C, Fitting Equations to Data with the Perfect Correlation Relationship. Hertfordshire Business School Working Paper (2015). Available at SSRN: https://ssrn.com/abstract=2707593 or http://dx.doi.org/10.2139/ssrn.2707593

**Funding:** This research received no external funding.

**Data Availability Statement:** No new data were created or analyzed in this study. Data sharing is not applicable to this article.

**Conflicts of Interest:** The author declares no conflicts of interest.